\documentclass[twocolumn,showpacs,preprintnumbers,amsmath,amssymb]{revtex4}
\usepackage{graphicx}
\usepackage{bm}
\usepackage{dcolumn}

\begin{document}


\title{Effect of rare events on out of equilibrium relaxation\\}

\author{Philippe Ribi\`ere, Patrick Richard,
Renaud Delannay and Daniel Bideau}
\affiliation{GMCM, UMR CNRS 6626, Universit\'e de Rennes I, campus de Beaulieu Bat 11A, F-35042 Rennes cedex, France}%

\author{Masahiro Toiya and Wolfgang Losert}
\affiliation{IREAP, Department of Physics, University of Maryland, College Park, Maryland 20742, USA}
\date{\today}

\begin{abstract}
This letter reports experimental and numerical results 
on particle dynamics in an out-of-equilibrium granular medium.
We observed two distinct types of grain motion:
the well known cage motion
, during which a grain is always surrounded by the same neighbors, and low 
probability jumps, during which 
a grain moves significantly more relative to the others.
These observations are similar to the results obtained for other
out-of-equilibrium systems (glasses, colloidal systems, etc.).
Although such jumps are extremely rare, by inhibiting
them in numerical simulations we demonstrate that they play a significant 
role in the relaxation of out-of-equilibrium systems. 

\end{abstract}

\pacs{45.70.-n,81.05}
\keywords{Granular compaction, cage effect, jump}
\maketitle

Glasses, colloids, and granular materials are characterized by structural disorder at the mesoscopic scale. These heterogeneous and metastable systems share a number of dynamical properties such as aging and slow relaxation, in which the cage motion of individual grains plays a significant role. Many experimental and numerical efforts have been made to observe and understand the out-of-equilibrium behavior exhibited by these systems~\cite{NatureMaterials,Chicago,Dauchot2005,Weeks2000,Bertin2005,Berthier2003,Pouliquen2003,Berg2002}.
A full understanding of the macroscopic behaviors of these systems requires 
detailed observations at the local scale (i.e. at the scale of a particle). 
The local dynamics in glasses is numerically simulated
with Lennard-Jones liquids and the
motion of atoms can be correlated to the $\alpha$ and $\beta$
relaxation regimes~\cite{Berthier2003}. 
Similar observations are obtained experimentally 
on colloidal systems using confocal microscopy~\cite{Weeks2000}, granular media in 2D~\cite{Dauchot2005},
or 3D granular assemblies immersed in  index matching fluid~\cite{Pouliquen2003}. 
For all theses systems, 
the cage effect is clearly identified.
 Here we present an extensive experimental
and numerical study 
of particle motion in 3D 
granular packing under gentle vertical tapping.
{Although this system is not thermal (its thermal energy is irrelevant in
comparison with the energy needed to move a macroscopic grain), it is often
presented as an ideal system to study out-of-equilibrium systems.
From a microscopic point of view, this analogy
is based on the idea that the geometry of the grains plays a major role, 
similar to  geometrical frustration  in thermal glasses. 
Indeed, like glasses, a 
granular assembly can be trapped in a metastable configuration 
unless an external
perturbation such as shear or vibration is applied.
Note that, unlike thermal energy, mechanical agitation
of grains is, in general, neither stochastic nor isotropic.
We expect, as others have found~\cite{Chicago,Philippe2002,Dauchot2005,Pouliquen2003,Danna2003,Josserand2000,Makse2002}, that the results obtained are valid for any of the
above mentioned out-of-equilibrium systems.}
Let us recall that when a dense granular sample is submitted to gentle
external mechanical perturbations, its packing fraction $\Phi$ increases quickly at the beginning, then gradually approaches an asymptotic value. This slow compaction is similar to the slow dynamics observed in other out-of-equilibrium systems~\cite{Chicago,Philippe2002,Danna2001}.

The experimental results presented in the first part of this Letter are obtained with $D = 0.6\mbox{ mm}$ diameter glass spheres placed in a glass cylinder of diameter $6\mbox{ mm}$. This container is placed on a plate connected to an electromagnetic exciter. The glass spheres are immersed in index matching oil containing a laser dye.  When imaged from the objective lens through the bottom of the sample container, beads appear as dark circles. Sequences of $10^4$ taps are carried out.
The only control parameter is the tapping intensity $\Gamma= \gamma_{max}/g$ where  $g = 9.81\mbox{ m.s}^{-2}$ and $\gamma_{max}$ is the maximum of the acceleration monitored by the accelerometer when the plate goes down~\cite{Ribiere2004}. 
It should be pointed out that we only used
taps for which the grains took off from the bottom
of the glass cylinder i.e. above the lift-off threshold.
Results from numerical simulations
are shown in the second part of this Letter.
In these simulations, the compaction of $4096$ monosize spheres of 
radius $R$ under tapping from a low packing state to a higher one is studied.
The container is a box with a flat $32R \times 32R$ square bottom with periodic boundary 
conditions imposed in the horizontal directions, and a free interface 
at the top of the packing. A tap is simulated in two stages. 
First, the packing is dilated vertically according to the 
law $(z-R)\longrightarrow (z-R)(1+\varepsilon)$, where $\varepsilon$ 
is the dilation parameter. This dilation increases 
the free volume and allows 
for the reorganization of beads. 
This linear behavior during a tap 
is also observed in our experiments and the parameter $\varepsilon$ 
is proportional to the square of 
the experimental parameter $\Gamma$~\cite{Ribiere2005}.
 The second stage simulates the gravitational redeposition, with a 
 nonsequential algorithm~\cite{Philippe2001}
 in order to observe the collective behavior. 
 The model only takes the steric constraints into account, which are fundamental for the compaction. While the simulation does not calculate the stress transmission in the packing, it describes relatively accurately the grain motion during a tap~\cite{Mehta1991}, contrary to other simulations such as the Tetris model for example, which show slow relaxation and aging, but whose dynamics have no direct relation with particle motion. Finally, our simulation results closely matches the experimental results. The experimental setup and the numerical simulation are described in detail in Refs~\cite{Philippe2002,Philippe2001}.\\
\begin{figure}
\centering
\includegraphics*[width=0.50\textwidth]{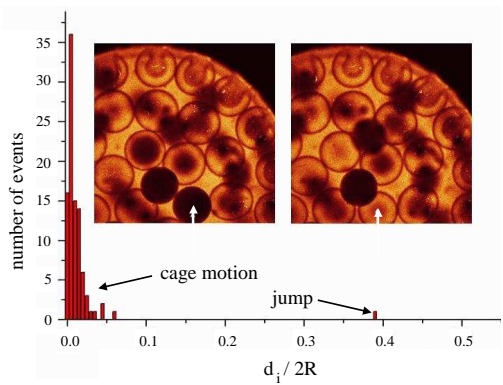} 
	\caption{(color online). Experimental observation of a jump, and distribution of bead displacements obtained by confocal imaging in an index matching fluid. The two pictures correspond to a horizontal cross section at the same altitude of the granular sample before and after the $126\mbox{th}$ tap of intensity $\Gamma=6$.
	 All the beads remain in their original layer except one bead (pointed by a white arrow) which falls from the upper to the lower layer. The distribution of bead displacements computed for 10 taps, five before, five after the $126^{th}$ tap, reveals the two scales of displacement.
	 }
	\label{fig:saut_exp}
\end{figure}
During an experiment in index matching fluid, cage motion and jumps 
 were observed.  
The cage motion is always characterized by the fact that a grain is 
always surrounded by the same neighbors.
A jump is a displacement during which a grain moves
significantly more than  the mean displacement of a 
grain. 
Such an event observed by confocal microscopy is shown in Fig.~\ref{fig:saut_exp}. The horizontal cross sections shown were taken during the same experiment, at the same height in the medium for two successive taps ($125\mbox{th}$ 
and $126\mbox{th}$).  On each picture, two different layers of the granular medium are visible.  The beads of the lower layer have a well-defined black outline and a bright central part, and the beads belonging to the upper layer are black. The vertical position of a bead is determined by measuring its apparent radius in the picture. Since the beads move only a small fraction of a bead diameter between taps, it is possible to track their positions from frame to frame.  A rare occurrence of jump is indicated by an arrow in ~\ref{fig:saut_exp}.  This bead, during a tap, has fallen from the upper visible layer to the lower one by a displacement larger than all the others. The amplitude of this jump between the two layers is $0.388\:D$ 
compared to the mean displacement in a layer, which is $0.00216\:D$, as shown in the distribution of bead displacement reported in Fig.~\ref{fig:saut_exp}.\\

\begin{figure}
\includegraphics*[width=0.50\textwidth]{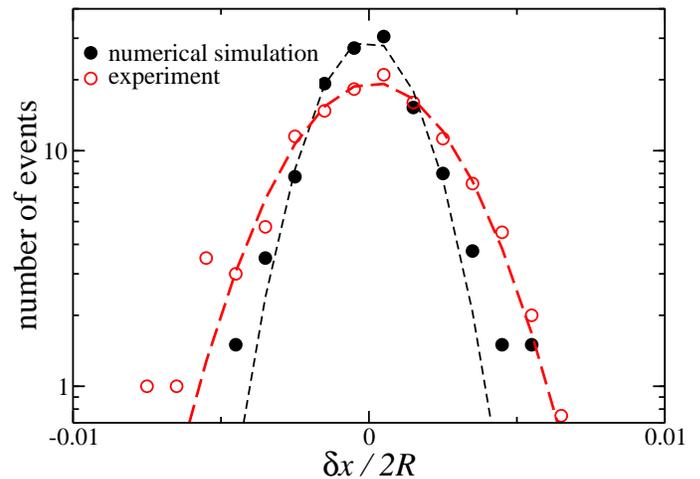}
	\caption{(color online). The histogram of displacements in a horizontal direction ($\delta x$) for our 3D experiment ($\Gamma=6$) and  for our corresponding numerical simulation ($\varepsilon=0.05$). The lines correspond to Gaussian
	fits. We observe deviations from the fits in the tails
	of the distributions.
	}
	\label{fig:histogram_deplacement}
\end{figure}

The study of the grain motion during our simulation confirms that the displacement of the beads exhibits two different scales. The smaller scale corresponds to the cage motion during which a grain explores the trap created by its neighbors. Previous results~\cite{Dauchot2005,Berthier2003} demonstrate the non-Gaussian character of the displacement distribution obtained in a 2D experiment and glass media, respectively. 
Such non-Gaussian behavior is also observed in our
experimental and numerical systems.
Figure~\ref{fig:histogram_deplacement} shows the histogram of 
displacements in the $x$ direction and the corresponding Gaussian fits.
Deviations from the fits can be observed in the tails of
the distributions. Note that the deviation is more important for the numerical
simulations.
The characteristics of the cage motion change during the simulation, because of the evolution of the packing fraction. Indeed, the average magnitude of this displacement decreases with the decrease of the void space in the medium.
%
\begin{figure}
	\centering
		\includegraphics*[width=0.50\textwidth]{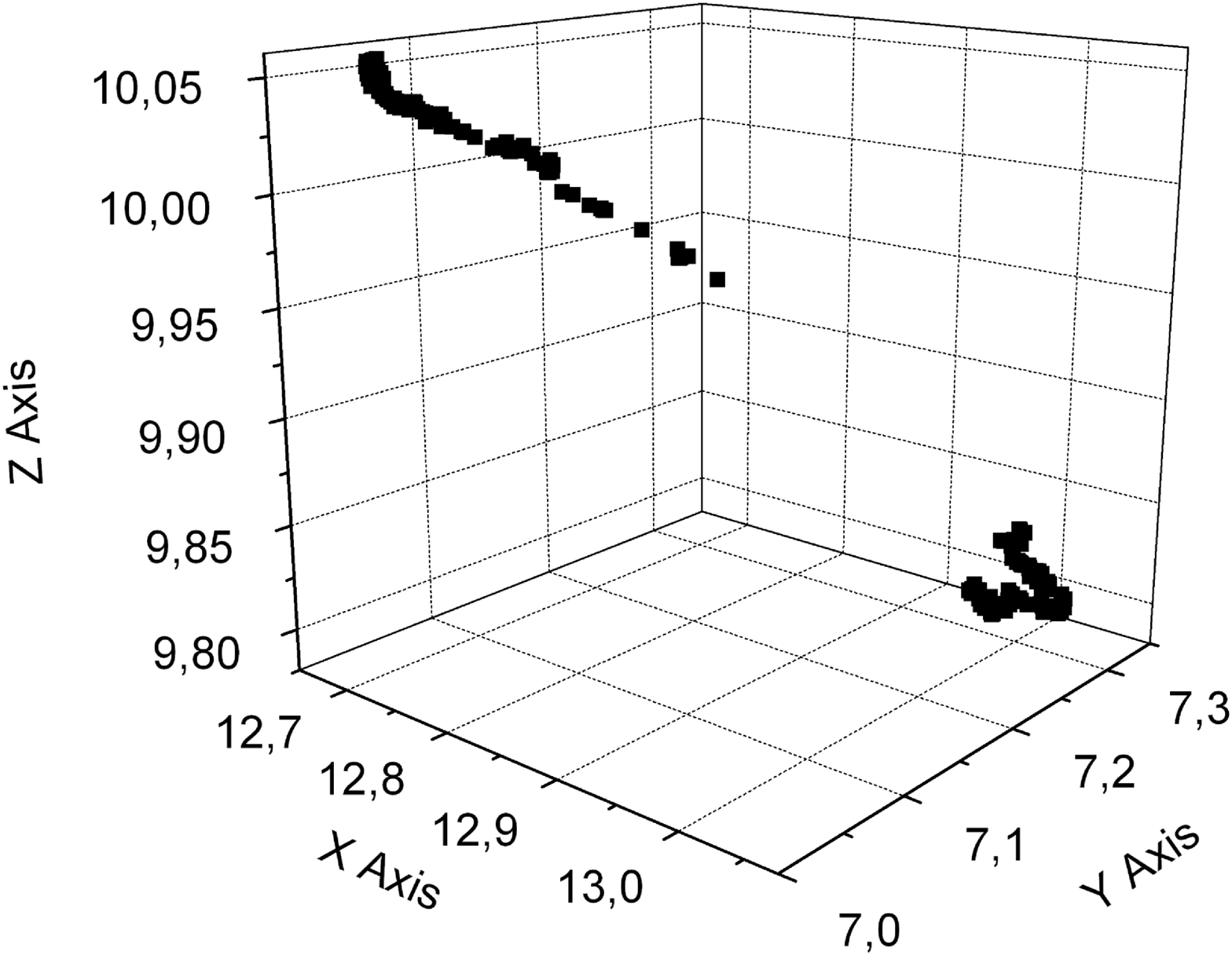}
	\caption{Trajectory of a bead, exhibiting one "jump", during the 
$100$ first taps. 
The position of the bead center is plotted after each tap during a simulation of linear dilation parameter $\varepsilon=0.05$. 
	Between the $46^{th}$ and the $47^{th}$ tap, a jump is visible corresponding to a large displacement. The units are in bead diameter.
	} 
	\label{fig:saut_simu}
\end{figure}
The larger scale corresponds to a jump: a very low probability event during which a bead enlarges its trap and thus, escapes from it. Figure ~\ref{fig:saut_simu} shows the trajectory of a bead which is subject to a "jump" during the $100$ first taps in the simulation. The initial drift in position is due to the increase of the packing fraction and, consequently to the decrease of the sample height. At the $46$th tap, the bead displacement is $0.48D$, although the average displacement (apart from the jump) is $0.0038D$.  We found that the jumps always yield a large displacement in the direction of gravity, as they are correlated to the compaction mechanism and correspond to the fall of a bead in a hole located under it. Note that the magnitudes of the cage motion and of the jumps 
are both 
of the same order in simulations and experiments. The simulation, thus, correctly represents both the local and global behaviors of the granular material. The jumps are extremely rare events, and therefore difficult to observe experimentally. In a simulation of $10\,000$ taps for $N=4096$ beads, only about $100$ jumps occurred, mainly at the beginning of the simulation, when the packing fraction is still far from the steady state value. Indeed a jump is possible only if the hole is large enough.
\begin{figure}
	\centering
		\includegraphics*[width=0.50\textwidth]{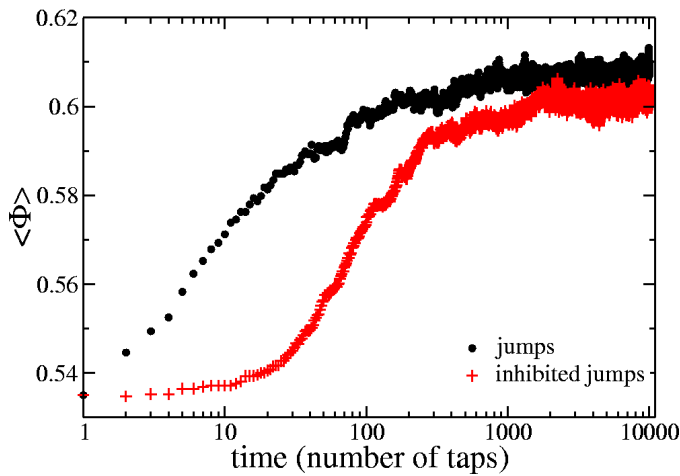}
	\caption{(color online). Evolution of the packing fraction with and without jumps 
	for the same dilation parameter $\varepsilon=0.05$.
	For the latter case,
	the increase of the packing fraction is slower at the beginning, as the number of inhibited events is greater at the beginning. Interestingly the packing fraction reached by the sample after a long time is lower without jumps than with jumps.}
	\label{fig:imp_saut}
\end{figure}
To understand the importance of the jumps, they are inhibited in the 
simulation, and the evolutions with normal jumps and with inhibited jumps 
are compared. To avoid jumps, the following test is added to the algorithm:  
If the vertical displacement of a bead is negative, i.e., if the bead moves down, and if its 
displacement $d_{i}$ is larger than twice the mean displacement 
of beads in the layer $\langle d\rangle$, this displacement is not 
accepted with probability 1. Instead, the probability of acceptance 
follows an exponentially decreasing function 
$\exp(-(d_i-2\langle d\rangle)/\delta)$. 
The brackets 
$\langle\rangle$ denote the average over a layer, and $\delta$ 
is calculated such that the probability is $10^{-6}$ 
when $d_{i}=R$ (the radius of a bead).
Note that we checked with a molecular dynamics simulation 
that prohibiting jumps does not lead to mechanically unstable packings.
The evolution of the packing fraction during the simulation with 
jumps and with strongly inhibited jumps is shown in Fig.~\ref{fig:imp_saut}. This graph reveals that, without significant jumps, the increase of packing fraction during the initial evolution is slower. This seems to make sense as some large displacements do not occur in this case. But, surprisingly, the 
packing obtained after a large number of taps is less dense without jumps than with jumps. In terms of energy landscape, this can be interpreted as two states that are separated by high potential barriers that cannot be circumnavigated 
by many slow events, and we can conclude that the energy landscape
exploration is completely different without jumps. Although jumps
are extremely rare events, they play a significant role in the evolution
of the media.
As we have noted earlier, the number of these events decreases 
as the system compacts,
and the great majority of jumps happens during the
first $250$ taps. So the initial behavior determines the subsequent evolution.
Somewhat similar to the well known memory effects 
{reported first in Ref~\cite{Josserand2000}},
the medium never
completely forgets its initial evolution.
\begin{figure}
\begin{center}
\includegraphics*[width=0.4\textwidth]{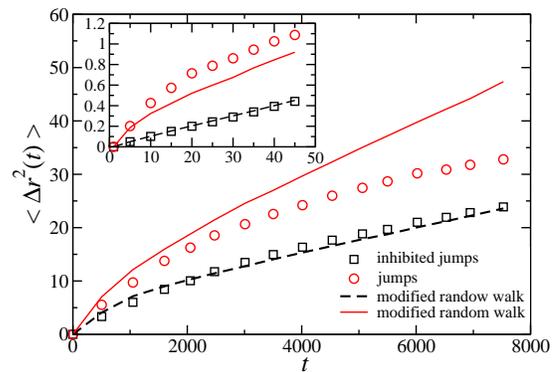} 
\caption{(color online). Evolution of 
$\langle \Delta r^2\rangle$
with $t$ for simulations with jumps and with inhibited jumps,
but for the same dilation parameter $\varepsilon=0.05$.
The random walk created 
correctly describes the behavior without jumps
but fails to reproduce the behavior observed with jumps.}\label{fig:mvt}
\end{center}
\end{figure}
\noindent To have better insight into the effect of jumps,
we study the evolution of
$\langle \Delta r^2(t)\rangle=\langle(\overrightarrow{r(t)}-\overrightarrow{r(0)})^2\rangle$,
 where $\overrightarrow{r(t)}$ is the grain position
 at time (i.e. number of taps) $t$,
with and without jumps. 
The brackets $\langle\rangle$ denote the average over the grains.
The results are
reported in Fig.~\ref{fig:mvt} for a dilation parameter $\varepsilon = 0.05$.
In the two cases a subdiffusive behavior is found. Note that it is not possible to compare directly the two evolutions, since the packing fractions
are  not the same with jumps or with inhibited jumps. A modified random walk
model that mimics the cage motion can, nevertheless, shed new light on the diffusive
properties without jumps, where cage motion is the only allowed displacement.
Two main characteristics were introduced in this model. First, an increase in packing fraction (thus decrease in cage size) results in decreased mean grain displacements.  Second, the high anticorrelation of the
motion inside a cage: Since the cage is a closed space the probability
to have at time $t$ a step in the opposite direction of the step at time $t-1$ is
higher than the probability to have these two steps in the same direction.
A 3D random walk is then generated for 4096 walkers. The 
step magnitude is
chosen to fit the decrease of the void in the sample : 
$d_i=\alpha(\Phi(t)^{-1}-1)^{1/3}$. The fitting parameter
$\alpha$ is adjusted to have the same mean displacement in the 
numerical simulation and in the random walk.
Finally, the direction of the step is chosen as follows:
The azimuthal angle $\theta(t)$ is chosen uniformly between $0$ and
$2\pi$
and the polar angle is given by $\varphi(t)=\pi-\varphi(t-1)+\eta$ ,
where $\eta$ is a 
Gaussian noise of average zero and of variance
$\pi(\Phi_{max}-\Phi)/(\Phi_{max}-\Phi_0)$.
{The values $\Phi_0$ and $\Phi_{max}$  are, respectively, the 
initial value of the packing fraction ($t=0$) and the packing 
fraction obtained at the end of the simulation (averaged between $t=9900$
and $t=10000$.)}
This decrease of the variance mimics the increase of anticorrelation with 
the increase in packing fraction $\Phi$.
We have reported in Fig.~\ref{fig:mvt} $\langle(\overrightarrow{r(t)}-\overrightarrow{r(0)})^2\rangle$ 
obtained
by the random walk model using the fit parameter $\alpha$ corresponding,
respectively, to the data with jumps and with inhibited jumps.
This random walk describes correctly the evolution of
$\langle(\overrightarrow{r(t)}-\overrightarrow{r(0)})^2\rangle$ without 
jumps, as expected, but fails to describe the behavior with jumps.
The initial evolution is too slow, as shown in the inset of Fig.~\ref{fig:mvt}, and the final evolution shows correlations which are too weak to describe the bead motion. The creation of a void large enough to allow a jump requires not only the motion of the first neighbors but also the motion of a large cluster~\cite{Losert2004}. Consequently, although the jumps are low probability events, they 
affect the motion of all the beads and thus the dynamics with jumps corresponds to more complex mechanisms.\\

We have studied experimentally and numerically the bead motion during 
compaction of a granular medium. We have demonstrated that extremely 
rare "microscopic" events (jumps) have huge consequences on the 
"macroscopic" behavior: Simulations show that
compaction is slower, and that the state reached after many taps
is less dense if the jumps are "inhibited".
This result is important and surprising because, 
in an equilibrium system, such rare events have negligible effects 
on its overall evolution.
{Such results, obtained for an out-of-equilibrium granular medium, should also be valid for other out-of-equilibrium systems
(glasses, collo\"\i ds, etc.) and could be useful for the study of segregation
of granular mixtures under shearing~\cite{Berton2003}.}\\

Acknowledgments:  We acknowledge funding from French ministry of
education and research (ACI \'energie et conception durable).
M.T. and W.L. acknowledge funding from NASA Grant NAG-32736, and equipment funding from the Office of Naval Research.

\end{document}